# PERFORMANCE ANALYSIS OF OFDM-BASED SYSTEM FOR VARIOUS CHANNELS

*I.Pramanik*[*], *M.A.F.M.Rashidul Hasan*[**], *Rubaiyat Yasmin*[**], *M. Sakir Hossain*[**], *Ahmed Kamal S.K.*[**],
[*]Patuakhali Science and Technology University, Bangladesh,
[**]University of Rajshahi, Bangladesh

**ABSTRACT:**

The demand for high-speed mobile wireless communications is rapidly growing. Orthogonal Frequency Division Multiplexing (OFDM) technology promises to be a key technique for achieving the high data capacity and spectral efficiency requirements for wireless communication systems ins the near future. This paper investigates the performance of OFDM-based system over static and non-static or fading channels. In order to investigate this, a simulation model has been created and implemented using MATLAB. A comparison has also been made between the performances of coherent and differential modulation scheme over static and fading channels. In the fading channels, it has been found that OFDM-based system's performance depends severely on Doppler shift which in turn depends on the velocity of user. It has been found that performance degrades as Doppler shift increase, as expected. This paper also performs a comparative study of OFDM-based system's performance on different fading channels and it has been found that it performs better over Rician channel, as expected and system performance improves as the value of Rician factor increases, as expected. As a last task, a coding technique, Gray Coding, has been used to improve system performance and it is found that it improves system performance by reducing BER about 25-32 percent.

**INTRODUCTION:**

Wireless communications is an emerging field, which has seen enormous growth in the last several years. The huge uptake rate of mobile phone technology, Wireless Local Area Networks (WLAN) and the exponential growth of the Internet have resulted in an increased demand for new methods of obtaining high capacity wireless networks.

Most WLAN systems currently use the IEEE802.11b standard, which provides a maximum data rate of 11 Mbps. Newer WLAN standards such as IEEE802.11a and HiperLAN2 [1], [2] are based on OFDM technology and provide a much higher data rate of 54 Mbps. However systems of the near future will require WLANs with data rate of greater than 100 Mbps, and so there is a need to further improve the spectral efficiency and data capacity of OFDM systems in WLAN applications. For cellular mobile applications, we will see in the near future a complete convergence of mobile phone technology, computing, Internet access, and potentially many multimedia applications such as video and high quality audio. In fact, some may argue that this convergence has already largely occurred, with the advent of being able to send and receive data using a notebook computer and a mobile phone. Although this is possible with current 2G (2[nd] Generation) Mobile phones, the data rates provided are very low (9.6 kbps – 14.4 kbps) and the cost is high (typically $0.20 - $1.30 AUD per minute) [3], [4], limiting the usefulness of such a service.

The goal of third and fourth generation mobile networks is to provide users with a high data rate, and to provide a wider range of services, such as voice communications, videophones, and high speed Internet access. The higher data rate of future mobile networks will be achieved by increasing the amount of spectrum allocated to the service and by improvements in the spectral efficiency. OFDM is a potential candidate for the physical layer of fourth generation mobile systems. This thesis presents techniques for improving the spectral efficiency of OFDM systems applied in WLAN and mobile networks.

**BASIC PRINCIPLES OF OFDM:**

Orthogonal Frequency Division Multiplexing (OFDM) is a multi-carrier transmission technique, which divides the available spectrum into many carriers. The information is modulated onto the sub-carrier by varying the phase, amplitude, or both. Each sub-carrier then combined to gather by using the inverse fast fourier transform to yield the time domain wave form that is to be transmitted. To obtain a high spectral efficiency the frequency response of each of the sub-carriers are overlapping and orthogonal. This orthogonality prevents interference between the sub carriers (ICI) and is preserved even when the signal passes through a multi-path channel by introducing a Cyclic Prefix, which prevents Inter-symbol Interference (ISI) on the carriers. This makes OFDM especially suited to wireless communications applications.

**SIMULATION RESULTS:**

An OFDM system has been modeled using Matlab to allow various parameters of the system to be varied and tested. The aim of doing the simulations is to measure the performance of OFDM under different channel conditions, and to allow for different OFDM configurations to be tested. The effect of different modulation techniques and modulation level on OFDM has also been tested. Moreover, different techniques have been employed to reduce BER (Bit Error Rate) rate. When OFDM performance has been tested in fading environment differential modulation scheme have been used instead of using one-tape equalizer. In the simulation data are assigned to all sub-carriers(i.e., There are no sub-carriers specifically assigned to pilots).

## 1. THE EFFECT OF GAUSSIAN NOISE ON OFDM-BASED SYSTEM PERFORMANCE:

The performance of OFDM system in the presence of White Gaussian noise has been tested using 16-QAM modulation scheme. Figure-1(a) shows the simulation result. It is seen from Fig-4.1 that when SNR (Signal to Noise Ratio) is low, the BER is very high and with the increase of SNR, BER reduces as expected. When SNR is low, the rate of reduction of BER is small but with the increase of SNR the rate of BER reduction increases rapidly.

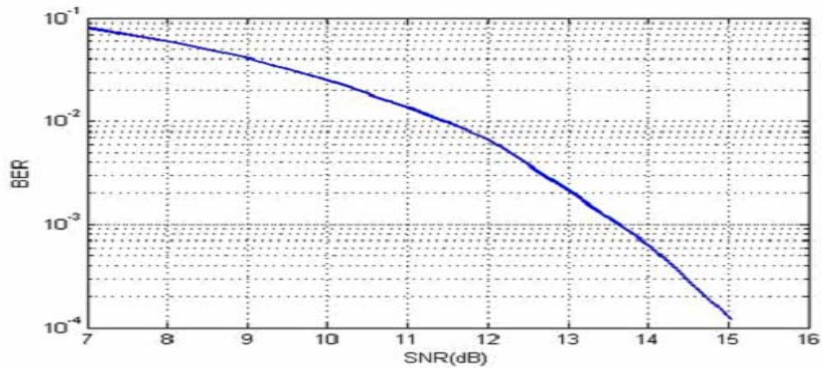

**Figure-1(a):BER versus SNR for OFDM using 16QAM**

This effect can be shown more easily with the help of constellation diagram. Fig-1(b) show constellation diagram for different SNR. It is seen from this figure that when SNR is low the constellation points corresponding to received signal are randomly located, thus it is difficult to recognize the actual location of each point in the constellation. This is due to channel noise. When SNR is increased to 11, about 70 percent points are located to their actual location, thus number of error reduces. When SNR is increased to 15, most of the points are located to their actual location, that is, number of bit error is very low. Thus simulated results corresponds to theory

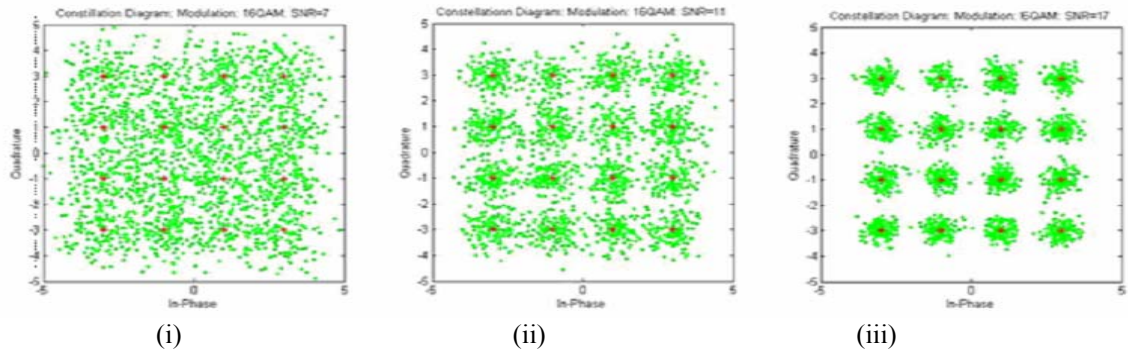

(i)  (ii)  (iii)
**Figure-1(b) Constellation Diagram for OFDM using 16OAM
(i) SNR:7   (ii) SNR:11   (iii) SNR:15**

## 2. EFFECT OF MODULATION TECHNIQUES ON OFDM-BASED SYSTEMS:

The effects of different modulation techniques and different modulation levels on OFDM-based system performance have also been investjgated in this paper. Fig-2 shows effect of modulation techniques on OFDM-based system. Here three modulation techniques, QAM, PSK, DPSK, are simulated. The modulation level is set at 16. It is seen from the fig-2 that lowest BER is achieved using QAM and BER is highest for DPSK. BER for using PSK is middle between other two. Let us now investigate the reasons for this result Since the phase difference in QAM is greater than that of PSK, so small variation of phase creates greater noise in PSK compared to QAM. The phase difference is increased in QAM by varying amplitude. ITU-T recommendation is to use three amplitude levels and twelve phase level for 16QAM. The OSI recommendation is to use four amplitude levels and eight phase levels. In our simulation we have used ITU-T recommendation. The BER for DPSK is the highest. The reason for this is that in DPSK, the output symbol phase corresponds to the phase difference between the present and previous symbols, and as a result the symbol noise is doubled (degrading the performance by 3 dB) compared with the phase noise of a single symbol (as used in coherent modulation like QAM, PSK). Thus in static channel, coherent modulation performs well than differential modulation.

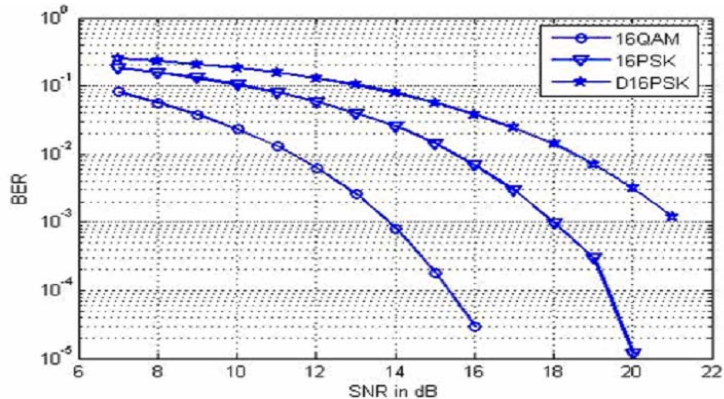

**Figure 2: Comparison of different modulation schemes for OFDM in static channel**

The effect of different modulation schemes on fading channels for OFDM is illustrated in Figure-3. Here simulation has been done using Doppler shift 5Hz and sampling period 2e-6 second. Here flat fading environment has been created. Figure shows that BER for DPSK is minimum in fading channel and BER of coherent modulation like QAM and PSK is very high. The reason for this will be investigated now. Phase rotations and amplitude scaling (important for QAM), greatly increase the error rate, or completely destroy all communications. This problem is however overcome by using channel equalization to remove this scaling of the channel before demodulation. The phase rotation of the channel and the amplitude scaling is measured using pilot symbols and pilot tones, which contain a known IQ transmission vector. Tracking of the channel requires continual updates in the channel equalization, thus regular pilot symbols/tones must be inserted into the transmission. The greater the number of pilot signals the faster the channel tracking rate, however this also causes significant overhead. But in our simulation no pilot symbols was inserted and no equalization was performed. Thus BER for using coherent modulation is very high.

**Performance of Coherent and Differential Modulation in Fading channel Modlevel:4**

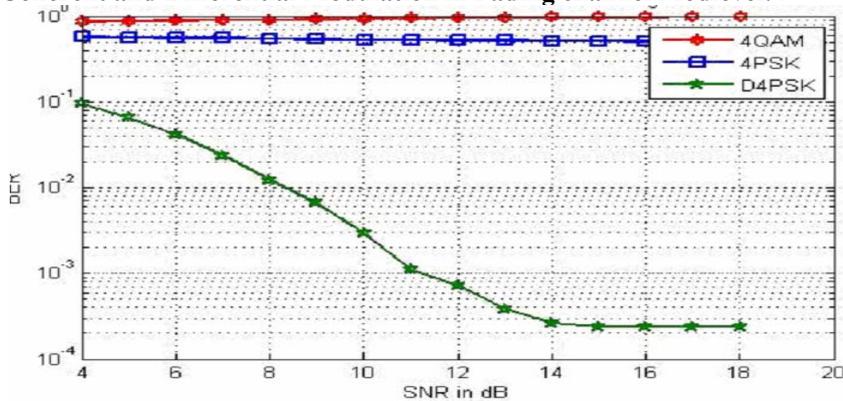

**Figure-3 Comparison of BER for OFDM for using different modulation schemes in fading channel (rayleigh channel here)**

On the other hand, differential modulation cancels out channel phase rotation, eliminating the need for additional channel equalization. Additionally the phase tracking of the channel is effectively updated at the symbol rate, thus tracking the channel very quickly. Differential modulation is thus highly suited to mobile communication.

A simulation has also been done to study the effect of modulation level on OFDM system performance. The simulated performance for the modulation schemes tested is shown in Figure- 4. These show the BER as a function of the Energy per Bit to Noise Ratio (Eb/No). This is a measure of the energy efficiency of a modulation scheme. If a higher Eb/No is needed to transfer data for a given modulation scheme, then it means that more energy is required for each bit transfer. Low spectral efficiency modulation schemes, such as BPSK and QPSK, require a lower Energy per Bit to Noise Ratio, and hence are more energy efficient. For a power limited system, with unbounded bandwidth, the maximum data rate could be achieved using BPSK or QPSK. However, in most applications the available bandwidth is the limiting factor and so the data rate is maximized by using a more spectrally efficient modulation schemes such as 256- QAM.

It is seen from figure-4(a) that BER increases as modulation level increases. The reason for this is that as modulation level increases amplitude and phase spacing between adjacent symbol decreases. Because of this, same noise produces greater error than that of smaller modulation level. The same result occurs for PSK where amplitude and frequency remains constant and only phase changes. As modulation level increases, phase spacing between adjacent symbol decreases which causes greater error.

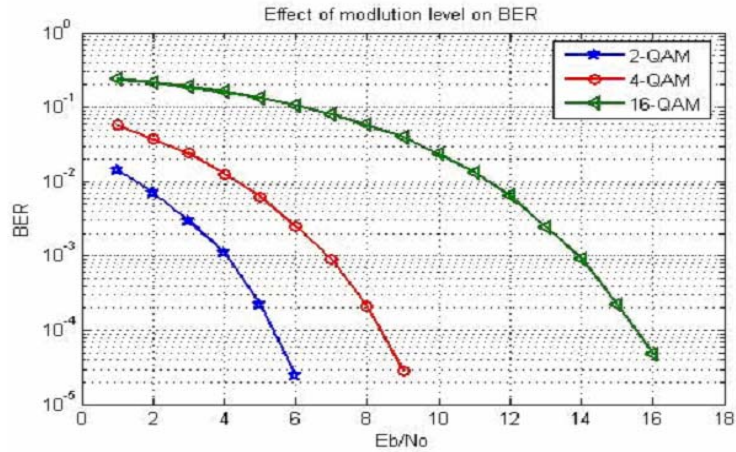

**(a) 2-OAM, 4-OAM, 16-OAM**

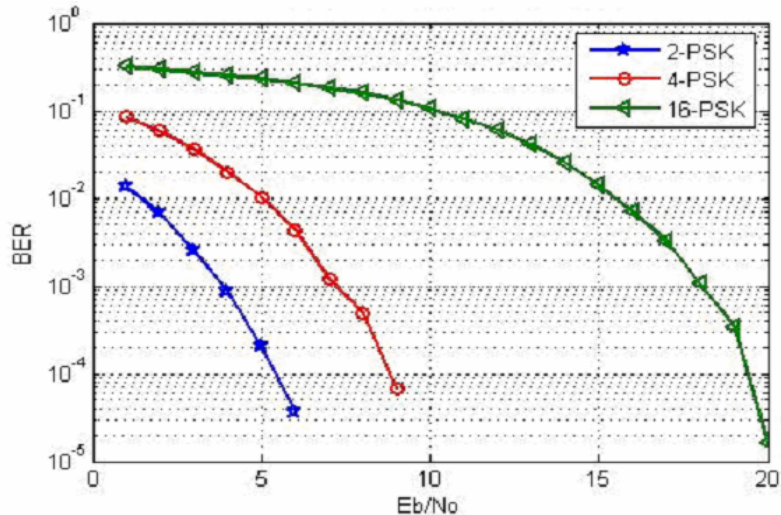

**(b) BPSK, QPSK, 16-PSK**

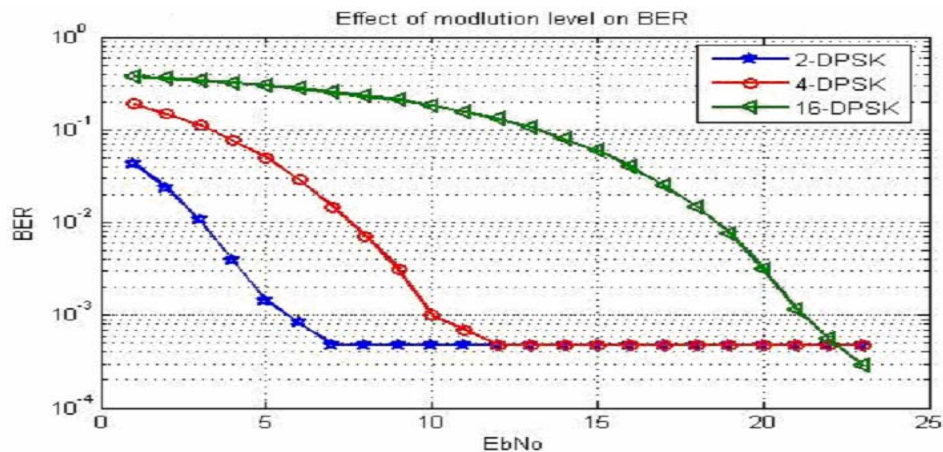

**(c) D-2PSK, D-4PSK, D-8PSK**

**Figure-4 BER versus the Energy per Bit to Noise Ratio for a selection of OFDM modulation schemes**

### 3. EFFECT OF FADING CHANNELS ON OFDM-BASED SYSTEMS:

The study of the effect of fading environment on OFDM system performance we have created slow-flat fading channel using MATLAB's build-in functions. To compensate error produced by flat fading, one of two method may be employed: either use differential modulation or use one tap equalizer [23]. In this simulation, differential modulation has been used. Let us first see how is OFDM performed in flat fading in case of Rayleigh fading channel. The simulation was done for pedestrian moving user who can cause.

Doppler Shift up to 5Hz for 900 MHz carrier frequency. Fig-5 shows the effect of Doppler shift, mobile user speed, on BER. The sampling period was set at 2e-6 second. It is seen from figure that as Doppler shift increases, that is mobile user approaches to the base station; Bit Error Rate increases as expected. The reason for this is that Doppler shift causes Doppler spread of the transmitted signal. Thus the bandwidth of the received signal get changed which makes it difficult for the frequency sensitive receiver to capture and recognize the transmitted signal accurately which causes increase in BER.

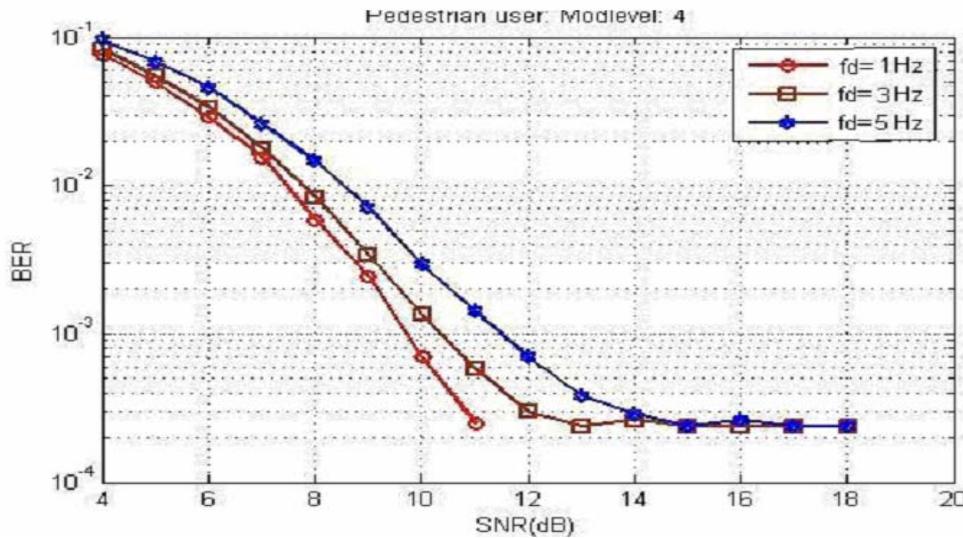

**Figure-5 Performance of OFDM in rayleigh fading channel for pedestrian user at different velocity**.

Now we shall compare the performance of OFDM in Rayleigh and Rician channel. Fig-6 shows such a simulation output. In this simulation D-4PSK was used as modulation scheme and Doppler shift and Rician factor were set at 40Hz and 5, respectively. From this figure it is seen that Bit Error Rate of OFDM over Rayleigh channel is far greater than that of Rician channel. The reason for this is very straightforward. From definition we know that this channel has no Line Of Sight(LOS) path and Rician channel has one LOS path or one major path. Since Rician channel has one LOS path so fading effect due to it must be lesser than that of Rayleigh channel. Thus Rayleigh channel produces greater BER than Rician channel.

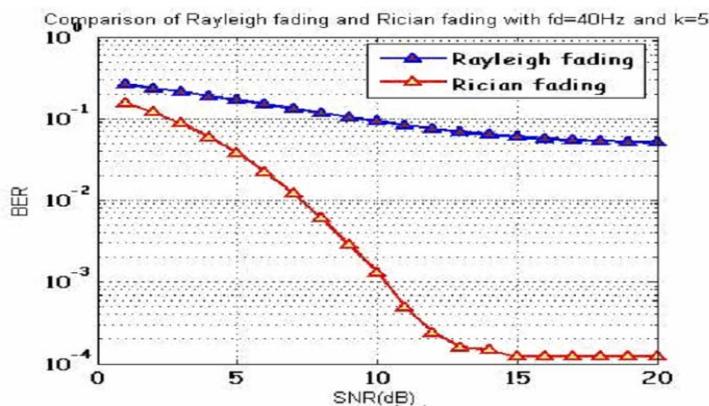

**Figure-6 OFDM over Rayleigh and Rician channel. Modulation:D-4PSK, Fd-40Hz, Rician factor-5**

Let us now compare the performance of OFDM in cases of Rayleigh channel, Rician channel and AWGN channel. Figure-7 depicts such a comparison. In this simulation, modulation level, Doppler shift and Rician factor k were set to 2, 40Hz and 5, respectively. DPSK was used as a modulation scheme. It is seen from

fig-7 that the BER for Rician channel lays between that of Rayleigh channel and AWGN channel. Why BER for Rician channel is less than Rayleigh channel has already been explained. Since AWGN has no fading effect and all kind of communication systems in any kind channel encounters White Gaussian noise (because of thermal noise, amplifier noise etc. in the receiver ) and Rician channel produces fading effect, though it is not severely, so Rician channel causes greater BER than AWGN channel. The performance of OFDM over Rician channel was also test through simulation in case of different Rician factor. Fig-7 also shows such performance comparison for different Rician factor. From definition of Rician factor, also called k factor, we know that it is the ratio of power in LOS or major path to the power in NLOS paths. So as the value of Rician factor increases the effect of multi-path effect decreases. Thus higher value of it indicates lower multi-path effect. That is, higher value of k-factor causes lower BER, and Fig-7 also justifies this phenomena.

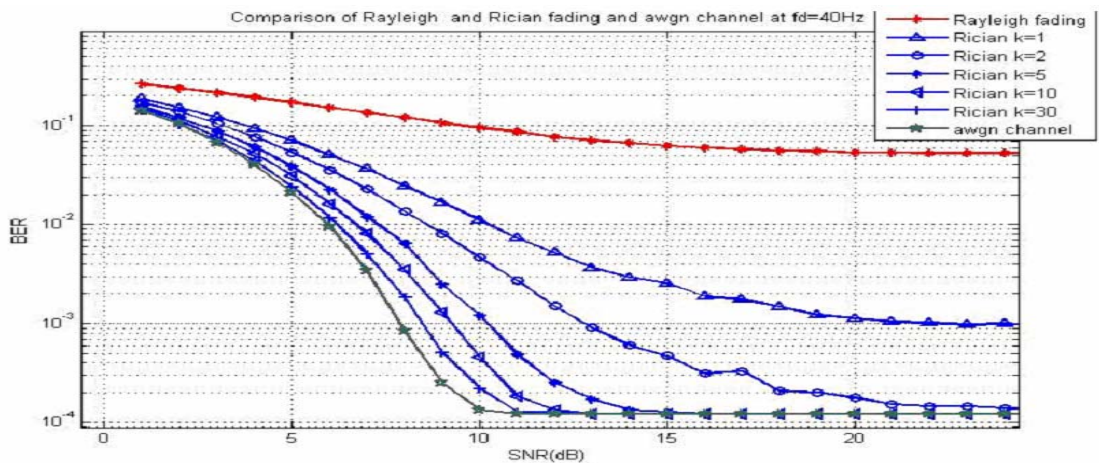

**Figure-7: BER for OFDM using D-2PSK in rician channel with k=1,2,5,10,30;Doppler shift:40Hz**

### 4. EFFECTS OF GRAY CODING ON OFDM-BASED SYSTEMS:

There are number of encoding techniques to reduce the Bit Error rate, that is, to improve ystem performance. Some examples of encoding techniques are Reed Solomon code, Convolution code, Gray code etc. In this project Gray Coding technique have been implemented to improve system performance. Fig-8 shows the effect of Gray coding on OFDM-based system performance for different channels. It is seen from Fig-8 that Gray Coding technique reduces BER which improves system performance significantly. It is observed during simulation that Gray Coding reduces BER about 20-30 percent. It is also seen that Gray Coding performance over different channels is almost same. Question may arise how Gray Coding technique improves system performance. The answer is that Gray Coding technique ensures that neighboring points in the constellation only differ by a single bit. This coding helps to minimize the overall Bit Error Rate as it reduces the chance of multiple bit errors occurring from a single symbol error.

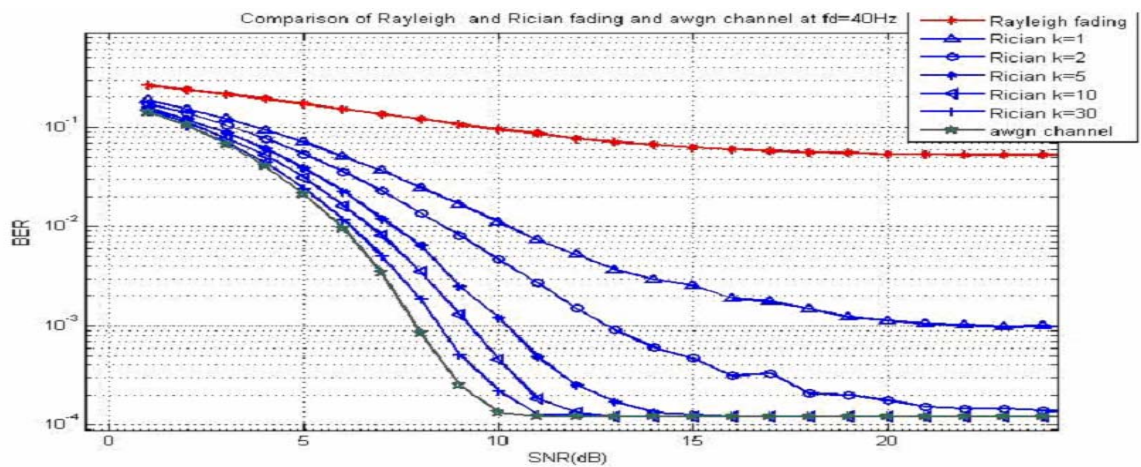

**Figure-8: OFDM performance improvement using Gray Coding Technique over various channels. Modulation:D-4PSK,fd=40Hz,k=5**

**CONCLUSION AND FUTURE WORKS:**

The current status of the research is that OFDM appears to be a suitable technique as a modulation technique for high performance wireless telecommunications. An OFDM link has been confirmed to work by using computer simulations, and some practical tests performed on a low bandwidth base-band signal. In this paper the performance of OFDM-based systems has been tested for various channels. Performance of OFDM-based systems found in this simulation is almost similar to theory.

Research can be done to illustrate the effect of the delay spread, peak power clipping etc. In this simulation no multi-path fading reduction techniques have been simulated and this is a field of future work also. One important major area which hasn't been investigated is the problems that may be uncounted when OFDM is used in a multi-user environment. One possible problem which may be encountered is the receiver may require a very large dynamic range in order to handle the large signal strength variation between users. Several modulation techniques for OFDM were investigated in this paper including BPSK, QPSK, 16PSK and 256PSK, however possible system performance gains may be possible by dynamically choosing the modulation technique based on the type of data being transmitted. More work could be done on investigating suitable techniques for doing this. OFDM promises to be a suitable modulation technique for high capacity wireless communications and will become more important in the future as wireless networks become more relied on.

**РЕЗЮМЕ**

У справжній статті досліджує ефективність OFDM-системи, заснованої на більш статичні і не статичний або затухання каналів. З метою розслідування цього, модель була створена і здійснюється з використанням MATLAB. Порівняння Крім того, було досягнуто між виступами послідовного і диференційованого схему модуляції більш статичної та затухання каналів. У згасання каналам, було встановлено, що OFDM-системи, заснованої на продуктивності сильно залежить від доплеровського зсуву, який, у свою чергу, залежить від швидкості руху користувача. Було встановлено, що погіршує показники, як доплеровській зсув збільшити, як і очікувалося. Цей документ також проводить порівняльне дослідження OFDM-системи, заснованої на діяльності на різних каналах зникає, і було встановлено, що вона виконує краще над Rician канал, як і очікувалося, і підвищує продуктивність системи, як вартість Rician фактора зростає, як і очікувалося. В якості останнього завдання методу кодування, Сірий кодування, була використана для підвищення продуктивності системи, і вона виявила, що вона підвищує продуктивність системи шляхом зниження BER близько 25-32 відсотків.

**РЕЗЮМЕ**

Данная статья исследует эффективность OFDM-системы, основанной на статических и не статических или затухающих каналах. Согласно результатам исследований, была разработана модель и выполнена симуляцияболее в среде MATLAB. Сравнение проводилось между когерентной производительностью и дифференциальной модуляционной схемой на статических и затухающих каналах. В затухающих каналах, было обнаружено, что производительность систем основанных на OFDM, сильно зависит от Доплеровского смещения, которое в свою очередь зависит от скорости объекта. Было обнаружено, что производительность уменьшается при увеличении смещения Доплера, как и ожидалось. Этот документ так же проводит сравнительный анализ производительности OFDM-систем в различных затухающих каналах и было обнаружено, что лучшая производительность достигается в Rician канале, как и ожидалось, производительность возрастает с ростом Rician фактора, как и ожидалось. В последней задаче была использована техника кодирования Gray Coding, для увеличения производительности системы и было обнаружено, что она увеличивает производительность системы за счет уменьшения BER на 25-32 процента.